\documentstyle[12pt]{article}

\def\nn{\nonumber \\}
\def\d{^\dagger}
\def\brav{\langle v|}

\advance\oddsidemargin by -\textwidth
\advance\oddsidemargin by -1in
\evensidemargin=\oddsidemargin
\begin{document}

$\mbox{ }$
\vspace{-3cm}
\begin{flushright}
\begin{tabular}{l}
{\bf KEK-TH-426 }\\
{\bf KEK preprint 426 }\\
December 1994
\end{tabular}
\end{flushright}

\baselineskip18pt
\vspace{2cm}
\begin{center}
\Large
{\baselineskip26pt \bf Temporal-gauge String Field \\
                       with Open Strings}
\end{center}
\vspace{1cm}
\begin{center}
\large
{\sc Tsuguo Mogami}
\end{center}
\normalsize
\begin{center}
%\begin{tabular}{l}
{\it Department of Physics, Kyoto University,}\\
{\it Kitashirakawa, Kyoto 606, Japan}\\
		{\it and} \\
{\it KEK Theory Group, Tsukuba, Ibaraki 305, Japan}\\
%\end{tabular}
\end{center}
\vspace{2cm}
\begin{center}
\normalsize
ABSTRACT
\end{center}
{\rightskip=2pc %3pc
\leftskip=2pc %3pc
\normalsize
A string field theory including open string fields is constructed in the
temporal gauge.  It consists of string interaction vertices similar to the
light-cone gauge string field theory.  A slight modification of the
definition of the time coordinate
is needed because of the existence of the open string end points.
\vglue 0.6cm}

\newpage
String field theory is considered to be the most promising way to describe
physics at the Planck scale nonperturbatively.  Recently a new kind of
string field theory was formulated \cite{IK}.  In such a string field
theory the nonperturbative aspects of noncritical strings, which are
described by matrix model thechniqtes in \cite{DS,FKN}, may be easily
deduced.  In this string field theory, points on world sheet are
parametrized by the geodesic distance \cite{KKMW} on it.  This ``geodesic
distance'' may be most easily understood on a dynamically triangulated
surface.  In the continuum theory, it may be understood as a kind of gauge
fixing which we call ``temporal gauge'' \cite{FIKN} or ``proper time
gauge'' \cite{Nakayama}.  It is also possible to introduce matter degrees
of freedom \cite{IK2,IIKMNS} which have $c=1-6/{m(m+1)}$.

So far, these string field theories have involved only closed string
fields.  In this article we will propose a way to introduce open string
fields into this ``temporal gauge'' string field theory for the case $c=0$.

Let us specify the model using via a matrix model.  Consider a matrix
model action
\begin{equation}
	S = \vec s^{\,*} \vec s  -  \mu \vec s ^{\,*} \phi \vec s
	+ {1 \over 2} {\rm tr} \phi^2
	- {\lambda \over 3} {\rm tr} \phi^3 .
\end{equation}
Here $\phi$ is a Hermitian matrix,and $\vec s$ is a vector which is also a
dynamical variable.
In this matrix model, open strings appear owing to the first two terms,
where we define $\mu$ to be the tension along the world line of an open string
end point.
It may be possible to obtain a string field theory by carefully examining the
continuum limit of the matrix model \cite{Watabiki}.
However, here we directly consider the continuum theory and fix the form of
the string field Hamiltonian from consistency conditions.

Let us recall the definition of the temporal gauge.  Consider a string
world sheet with a Euclidean metric.  The ``time'' coordinate of a point on
the surface may be defined as the minimum distrance from the initial
boundaries, where distance is measured by using the metric on the world
sheet.  As this ``time'' goes by, strings keep splitting and joining to
form world sheets.  It should also be noted that disappearance of an
infinitesimal string also occurs in this gauge.  Each string is
parametrized only by its length.  In order to describe such processes as
splitting and joining, it is convenient to introduce operators which
represent annihilation and creation of a string.
\begin{eqnarray}
	[ \Psi(l),\Psi^\dagger(l') ] = \delta(l-l'),	\nn{}
	[ \Phi(l),\Phi^\dagger(l') ] = \delta(l-l').
\end{eqnarray}
Here $\Psi(l)$ is the operator which annihilates an open string with
length $l$ and $\Psi^\dagger(l)$ creates an open string with length $l$.
$\Phi^\dagger(l)$ creates a closed string with a marked point. This convention
takes care of the statistical weights.
Of course, other commutators vanish.

The bra and ket vacua ($\langle 0|$ and $|0\rangle$) satisfy
\begin{eqnarray}
	\Psi(l) |0\rangle =0 ,~~~~~~~ \langle0| \Psi^\dagger(l) =0 ,	\nn
	\Phi(l) |0\rangle =0 ,~~~~~~~ \langle0| \Phi^\dagger(l) =0 ,
\end{eqnarray}
for all $l$.

The amplitudes with $i$ open strings and $j$ closed strings might be
calculated in this formalism as
\begin{equation}
	\lim_{D \to \infty} \langle 0| {\rm e}^{-D H}
		\Psi\d(l_1) \cdots \Psi\d(l_i)
		\Phi\d(l_{i+1}) \cdots \Phi\d(l_{i+j}) |0 \rangle,
			\label{amplitude}
\end{equation}
using the string field Hamiltonian $H$, which describes the time evolution. It
is the peculiarity of the Euclidean theory, that all the strings eventually
disappear.  This is why we take the limit $D \to \infty$.

The Hamiltonian which we propose here is
\begin{eqnarray}
H &=& \int dl_1\, dl_2\: \Phi\d(l_1) \Phi\d(l_2) \Phi(l_1+l_2)\, (l_1+l_2)\nn
  &+& \int dl_1\, dl_2\: \Psi\d(l_1) \Psi\d(l_2) \Psi(l_1+l_2)		\nn
  &+& 2\int dl_1\, dl_2\: \Psi\d(l_1) \Phi\d(l_2) \Psi(l_1+l_2)\: l_1	\nn
  &+& g \int dl_1\, dl_2\: \Phi\d(l_1+l_2) \Phi(l_1) \Phi(l_2)\, l_1 l_2
  	\label{z1} \\
  &+& 2g\int dl_1\, dl_2\: \Psi\d(l_1+l_2) \Phi(l_1) \Psi(l_2)\, l_1 l_2\nn
  &+& \sqrt g \int dl_1\, dl_2\, dl_3\, dl_4\:
	\Psi\d(l_1+l_3) \Psi\d(l_2+l_4) \Psi(l_1+l_2) \Psi(l_3+l_4)	\nn
  &+& \sqrt g \int dl\: \Psi\d(l) \Phi(l)\, l	\nn
  &+& \int dl\: \rho(l) \Phi(l)	+ \int dl\: \eta(l) \Psi(l)		\nn
  &+& 2\sqrt g \int dl_1\,dl_2\: \Psi(l_1)\delta(l_1+l_2)\Psi(l_2) \nn
  &+& a \biggl\{ \int dl\: \Phi\d(l) \Psi(l)
  + \int dl\: \Psi\d(l) ( -{\partial \over \partial l} - m) \Psi(l) \nn
  &&+ \sqrt g \int dl_1\, dl_2\: \Psi\d(l_1+l_2)\Psi(l_1)\Psi(l_2) \biggr\}
  	\nonumber
\end{eqnarray}
All the length variables are integrated from infinitesimal minus value to
infinity
to make
terms with singularity at zero length well defined.
This Hamiltonian is derived in the following way.

Considering the processes which might occur during the infinitesimal
``proper time'' evolution, only terms appearing on the right hand side of
eq(\ref{z1}) are possible for the Hamiltonian $H$.  The statistical weights
in the integrals ( e.g.  $l_1 l_2$ in the fourth term of eq.(5) ) are
determined in the same way as in the closed string case.  $g$ is the string
coupling constant.  The power of $g$ in front of each term is determined by
considering how the topology of the surface will change in each process.
We know $g$ has dimension $5$ in mass dimension for $c=0$ theory.  The
operator $\Phi\d(l)$ has dimension $5/2$ and $\Psi\d(l)$ has dimension
$3/2$.  We can see all the terms above have dimension $1/2$ except for the
terms in the brace.  Terms in the brace have dimension $1$, where the
multiplication factor $a$ is arbitrary and has dimension $-1/2$.  We label
``tadpole terms'' those terms which have no creation operator.  These terms
represent the disappearance of a string with infinitesimal length and are
the peculiarity of this gauge.  We will discuss these terms later and fix
the unknown functions $\rho(l)$ and $\eta(l)$.  Up to this point, we do not
know with what numerical factors all the terms in eq.(\ref{z1}) should
appear in the Hamiltonian.  The following consistency conditions will fix
these factors.

In the string filed theory of closed stings \cite{IK2},
the Hamiltonian was written as
\begin{equation}
	H_c = \int_0^\infty dl\: T_c(l) \Phi(l).
\end{equation}
Here the currents $T_c(l)$ satisfy the Virasoro algebra: $[T_c(l),T_c(l')] = g
 ll' {(l'-l)/(l'+l)} T_c(l+l')$. The fact that $T_c(l)$ forms a closed algebra
is essential to the consistency of the theory. It means that the equation
\begin{equation}
	\langle v_c| T_c(l) = 0. \label{eq7.5}
\end{equation}
is integrable.  Here $\langle v_c| \equiv \lim_{D \to \infty} \langle 0|
{\rm e}^{-D H_c} $.  Eq.(\ref{eq7.5}) is equivalent to the SD equations of
the string field theory, which are usually written in the form of $\langle
v_c| T(l) |{\rm anystate}\rangle = 0$.  One can show that the integrability
of eq.(\ref{eq7.5}) is closely related to the residual general coordinate
invariance in this gauge \cite{IK2}.  We expect a similar situation in our
Hamiltonian.  If we include open string fields, the Hamiltonian will appear as
\begin{eqnarray}
	H = \int_0^\infty dl\: T_1(l) \Phi(l)
		+ \int_0^\infty dl\: T_2(l) \Phi(l) + \cdots 	\nn
	+\int_0^\infty dl\: J_1(l) \Psi(l)
		+ \int_0^\infty dl\: J_2(l) \Psi(l) + \cdots.
\end{eqnarray}
The SD equations may be $\langle v| T_i(l) = \langle v| J_i(l) =0 $, where
$\langle v| = \lim_{D \to \infty} {\rm e}^{-D H}$.  In this case, it is
impossible to form a closed algebra because of the four point vertex of
open strings appearing in (\ref{z1}).  We must relax this requirement and
require that $\langle v| [T_i(l),T_j(l')] = \brav [T_i(l),J_j(l')] = \brav
[J_i(l),J_j(l')] = 0$ can be deduced from $\brav T_i(l) = \brav J_i(l) =
0$.  This consistency will fix the numerical factors in $T$'s and $J$'s.

For the time being, we disregard the tadpole terms. The following three
currents
form a closed system of constraints.
\begin{eqnarray}
T(l) &=& \int_0^l dl_1 \Phi\d(l-l_1) \Phi\d(l_1)
	+ g \int_0^\infty dl_1 \Phi\d(l+l_1) \Phi(l_1)		\nn
	&&+ g \int_0^\infty dl_1 \Psi\d(l+l_1) \Psi(l_1)
	+ \sqrt g \Psi\d(l),						\\
J(l) &=& \sqrt g \int dl_1\, dl_2\, dl_3\, dl_4\:
	\Psi\d(l_1+l_3) \Psi\d(l_2+l_4) \Psi(l_1+l_2) \delta(l_3+l_4-l)\nn
	&& +2\int_0^l dl_1\: \Psi\d(l-l_1) \Phi\d(l_1)\: (l-l_1)
	   +g\int_0^\infty dl_1\: \Psi\d(l+l_1) \Phi(l_1)\: l_1 l	\nn
	&& +\int_0^l dl_1\: \Psi\d(l-l_1) \Psi\d(l_1)		\\
J'(l)&=& \Phi\d(l) + {\partial \over \partial l} \Psi\d(l)
	+ \sqrt g \int_0^\infty dl_1 \Psi\d(l+l_1) \Psi(l_1).
\end{eqnarray}
All the terms included in these currents are obtained by stripping off one
of the annihilation operators from terms in eq.(\ref{z1}).
(Furthermore,
$T(l)$ is divided by $l$.)  We see each current put on the right of
$\langle v|$ corresponds to the Schwinger-Dyson equation of the matrix model.
To calculate commutators of these currents, it is easier to work with the
following current.
\begin{eqnarray}
J_3(l) &=& \biggl\{ J(l)-\int_0^\infty dx
		(2x-l)\: J'(l-x) \Psi\d(l) \biggr\}/l,			\nn
	&=&\sqrt g\int_l^\infty dx_1 \int_0^\infty dx_2 \Psi\d(x_1)\Psi\d(x_2)
		\Psi(x_1+x_2-l)						\\
	& & +\int_0^l dl_1\: \Psi\d(l - l_1) \Phi\d(l_1)
		+g\int_0^\infty dl_1\: \Psi\d(l+l_1) \Phi(l_1) l_1
		+ \Psi\d(l) \Psi\d(0).				\nonumber
\end{eqnarray}
Using this current the integrability of the constraints becomes
\begin{eqnarray}
	[T(l),T(l')] 	&=& g\: (l'-l)\: T(l+l'),	\label{TT}	\\{}
	[J_3(l),J_3(l')]&=& \sqrt g \int_l^{l'}dx\:J_3(l+l'-x)\Psi\d(x) ,\\{}
	[J_3(l),T(l')]	&=& g\: (l'-l) J_3(l+l')			\nn{}
	    &&+\sqrt g \biggl\{ \int_l^{l+l'}dx\:(x-l)+\int_0^{l'}dx\:(x-l')
		\biggr\} J'(l+l'-x) \Psi\d(x),				\\{}
	[J_3(l),J'(l')]	&=& -\sqrt g J_3(l+l')
		+\sqrt g\int_0^{l'} dx\: J'(l+l'-x) \Psi\d(x),		\\{}
	[T(l),J'(l')]	&=& g l' J'(l+l'),				\\{}
	[J'(l),J'(l')]	&=& 0.			\label{JJ}
\end{eqnarray}
In doing the calculation, it is seen that numerical factors are fixed
leaving only the freedom to choose the string couping $g$.  ( In fact
eq(\ref{fg}) fixes the other freedom to rescale $\Psi(l)$.  )

Now we must include the tadpole terms since the above constraints do not
reproduce the tree level amplitude.  To the Hamiltonian $H$ we add the
tadpole term for a closed string $\int dl \rho(l) \Phi(l)$.  The constraint
now becomes $\langle v| \{ l\,T(l) + \rho(l) \} = 0$.  This eqation gives
$\langle v| \{ l\,T(l) + \rho(l) \} |0\rangle = 0$.  Taking the zeroth
terms in $g$ we obtain an equation relating $\rho(l)$ and the disk
amplitude $f(l)$.  We know the result of the matrix model \cite{BIPZ} for
$f(l)$.  In the Laplace transformed form it becomes:  $\tilde f(\zeta) =
(\zeta -\sqrt t/2)\sqrt{\zeta+\sqrt t}$, where $t$ is the cosmological
constant.  This result with the constraint equation above fixes the
function $\rho(l)$ \cite{IK} as
\begin{equation}
	\rho(l) = 3\,\delta''(l) - {3 \over 4}t\, \delta(l).
\end{equation}
The equation (\ref{TT}) should be modified to:
\begin{equation}
	[l\:T(l)+\rho(l),l' T(l')+\rho(l')]
		= g ll'{l'-l \over l+l'} \{ (l+l')\:T(l+l') + \rho(l+l') \}.
			\label{TTr}
\end{equation}
In order for eq.(\ref{TTr}) to be valid,
$ll'{l'-l \over l+l'} \rho(l+l')$ on the right hand side of eq.(\ref{TTr})
should vanish.
This term seems rather ill-defined because of the singularity at $l+l'=0$.
Here we deal with the singularity by Laplace transforming it with
respect to both $l$ and $l'$.
\begin{equation}
	\int_{-\epsilon}^\infty dl\: {\rm e}^{-l\zeta}
	\int_{-\epsilon}^\infty dl' {\rm e}^{-l'\zeta}
	ll'{l'-l \over l+l'} \{3\,\delta''(l) - {3 \over 4}t\,\delta(l) \} =0.
		\label{eq21}
\end{equation}
Here the limit $ \epsilon \to 0$ is taken after integration.
This equality is essentially related to the integrability and the general
covariance
of this theory \cite{IK2}.

The tadpole terms are needed also for open strings.
The tree level one open string amplitude $h(l)$ may be calculated from the
disk amplitude for a closed string $f(l)$ as,
\begin{equation}
	h(l) = \int_0^\infty dl'\: {\rm e}^{-l' m} f(l+l') .	\label{fg}
\end{equation}
Here $m$ is the tension along the world line of open string end points.
The condition $\langle v| J(l) |0 \rangle = 0$ should be modified to
include tadpoles.  We must add $-m\Psi\d(l)$ to $J'(l)$ and to $J(l)$
\begin{equation}
	\eta(l)
	+2\sqrt g \int_{-\epsilon}^\infty dl\: \delta(l+l_1) \Psi(l_1),
\end{equation}
where $\eta(l)$ is fixed as $-2\delta'(l) -m\delta(l)$.
$\int dl\: \delta(l+l_1) \Psi(l_1)$ seems to be zero in Laplace transformed
form, but this does not hold if $\Psi(l)$ is singular at $l=0$. In fact this
term
survives in the commutator $[J(l),J'(l')]$.
Using eq.(\ref{TT})-(\ref{JJ}), it is easily checked
that the equations
\begin{eqnarray}
	\langle v| \{ l\,T(l) + \rho(l) \}  = 0, \\
	\langle v| \biggl\{ J(l) + \eta(l)
	    +2\sqrt g \int dl\: \delta(l+l_1) \Psi(l_1) \biggr\} = 0,	\\
	\langle v| \{J'(l) -m\Psi\d(l) \} = 0,
\end{eqnarray}
are integrable using formulas such as eq.(\ref{eq21}).

So the Hamiltonian we are seeking is
\begin{eqnarray}
	H &=& \int_0^\infty dl\: \{ l T(l) + \rho(l) \} \Phi(l)	\nn
		&&+\int_0^\infty dl\: \biggl\{ J(l) + \eta(l)
		+ 2\sqrt g \int dl\: \delta(l+l_1) \Psi(l_1) \biggr\} \Psi(l)
			\label{H} 	\\
		&&+a \int_0^\infty dl\: \{J'(l) -m\Psi\d(l) \} \Psi(l).
				\nonumber
\end{eqnarray}
We can see eq.(\ref{H}) is the same thing as eq.(\ref{z1}).  A few comments
about this Hamiltonian are in order.  It has ambiguities coming from the
overall factors of the three currents.  The ratio of the first term to the
second term in eq(\ref{H}) must be 1, because $\int dl_1\, dl_2\:
\Phi\d(l_1) \Phi\d(l_2) \Phi(l_1+l_2)\, (l_1+l_2)$ in the first term
represents the same process as $\int dl_1\, dl_2\, l_1\:  \Psi\d(l_1)
\Phi\d(l_2) \Psi(l_1+l_2)$ at the point where the string splits.  The
remaining ambiguities are absorbed into rescaling of $a$ and ``time'' $D$.
The constant $a$ has dimension $1/2$, so it should not be fixed.  The term
with $a$ describes proper time evolution along the open string boundaries.
Therefore, if we change $a$, we obtain a slightly different definition of
our time coordinate.  This difference will be seen near the world lines of
open string end-points.  If $a \neq 0$, $H$ looks like the light-cone gauge
string field theory of Kaku-Kikkawa, since $H$ has both the vertex in which
an open sting splits and a vertex in which two open strings merge.
However, $a=0$ might be the most natural choice because the world sheets
have fractal geometry, and the geodesic curves through the surface will be
shorter than paths along the boundary.  The naive continuum limit of the
stochastic quantization of the matrix model as in \cite{JR} yields $a=0$.

In this article, open string fields are introduced into the string field
theory in the temporal gauge.  The form of the Hamiltonian was fixed by
considering possible processes and consistency.  The string vertices
appearing in this formalism look like the vertices of the light-cone gauge
string field theory.

\section*{Acknowledgements}
The author is grateful to H. Kawai, N. Ishibashi and N. Sasakura for useful
discussions and suggestions.

%\newpage

\vfill
\end{document}